\documentclass[reprint,onecolumn,amsmath,amssymb,prd]{revtex4-2}
\usepackage{graphicx}
\usepackage{dcolumn}
\usepackage{bm}
\usepackage{hyperref}
\usepackage{graphics}
\usepackage{slashed}
\usepackage{amssymb}
\usepackage{natbib}
\usepackage{amsmath}
\usepackage{amsthm}
\usepackage{url}
\usepackage[usenames,dvipsnames,svgnames,table]{xcolor}
\usepackage[T1]{fontenc}

\newcommand{\fracc}[2]{\frac{\textstyle{#1}}{\textstyle{#2}}}

\makeatletter
\newcommand{\tpitchfork}{%
  \vbox{
    \baselineskip\z@skip
    \lineskip-.52ex
    \lineskiplimit\maxdimen
    \m@th
    \ialign{##\crcr\hidewidth\smash{$-$}\hidewidth\crcr$\pitchfork$\crcr}
  }%
}
\makeatother

\usepackage{xcolor}

\begin{document}

\title{Photon traps in nonlinear electrodynamics}

\author{\'Erico Goulart}
 \email{egoulart@ufsj.edu.br}
\affiliation{Federal University of S\~ao Jo\~ao d'El Rei, C.A.P. Rod.: MG 443, KM 7, CEP-36420-000, Ouro Branco, MG, Brazil
}

\author{Eduardo Bittencourt}
 \email{bittencourt@unifei.edu.br}
\affiliation{Federal University of Itajub\' a, BPS Avenue, 1303, Itajub\'a/MG - Brazil
}

\date{\today}

\begin{abstract}
We demonstrate the existence of photon traps within the framework of nonlinear electrodynamics. The trapping mechanism is based on the fact that, for null background fields, the optical metric reduces to the Kerr-Schild form, which plays a prominent role in the context of black hole physics. We then construct explicit examples where photons are confined in a region of spacetime, such that a distant observer cannot interact with them. Finally, we argue that the trapping scheme is quite universal, being entirely compatible with causality, energy conditions, and hyperbolicity.

\end{abstract}

\maketitle

\section{\label{sec:level1} INTRODUCTION}

Theoretical physicists have good reasons to investigate nonlinear extensions of Maxwell’s electrodynamics (NLED). In quantum field theory, the
polarization of the vacuum leads naturally to nonlinear corrections, which Euler-Heisenberg’s Lagrangian describes \cite{Euler1936}. In some material media, such as dielectrics and crystals, the complex interaction between the molecules and external electromagnetic fields can be described by an effective nonlinear theory, which is typically observed at very high light intensities such as those provided by pulsed lasers \cite{Born1999}. Heuristic nonlinear models have also been studied in the context of black hole solutions \cite{AYONBEATO1999,novello2002artificial,barcelo2005,breton2007nonlinear,Bittencourt2014,Bronnikov2018,PhysRevA.104.043523}, to create cosmological models that might avoid singularities \cite{NOVELLO2008}, to mimic dark energy \cite{PhysRevD.81.065026,benaoum2023}, and generate an accelerated expansion of the universe \cite{Novello2004,Kruglov2015}. Other important applications include bosonic open string theory where a nonlinear action naturally arises as an effective low-energy action and the so-called ModMax electrodynamics which share some symmetries with the linear theory \cite{Sorokin_2022,alam2022review,DUNNE_2005}.

In this paper, we focus on the possibility of creating photon traps within the framework of NLED. More generally, the idea of trapping photons and somehow confining them in a region is an active research area and is important for a plethora of fundamental reasons and practical applications. Indeed, several experimental mechanisms to slow down, localize, trap, and store photons in atomic media or photonic structures have been proposed in recent years (see e.g. \cite{koshelev2020subwavelength,hsu2013observation}). In order to demonstrate the existence of such traps in NLED, we start by interpreting the photon as a linear perturbation over a background field configuration, and relying upon the effective geometry technique, here called the optical metric. As is well known, light rays in a nonlinear theory propagate as null geodesics with respect to the optical metric, very much in the same way as they do in the context of general relativity. Then, we restrict the analysis to the case of null backgrounds, namely, we consider solutions such that both field invariants vanish: electric and magnetic fields are perpendicular and have the same magnitudes. After some algebraic manipulations, it is shown that the optical metric reduces to the Kerr-Schild form, where the corresponding null vector is a principal null direction of the electromagnetic energy-momentum tensor. This is a crucial intermediate point because the particular form of the optical metric permits us to apply well-known theorems of general relativity to the case of propagating photons in NLED. Finally, we move on to some interesting applications of the trappings.

Our construction is based on the fact that for null invariants, solutions of NLED are exactly the same as Maxwell's linear theory. Also, we choose some physically appealing null solutions that lead to optical metrics with special features when compared to the Minkowskian causal structure (see Ref.\ \cite{GoulartdeOliveiraCosta:2009pr} for a detailed classification of light cones in NLED). In particular, we shall see that both cones intersect along the principal direction of the electromagnetic field. However, depending on the field intensity, the optical cone tends to close toward the direction of the Poynting vector. In particular, there are critical situations where a given null optical direction is tangent to Minkowski time. Roughly speaking, this means that photons emitted in that direction would be at rest with respect to laboratory observers. Clearly, this phenomenon mimics what we know as an event horizon in black hole physics. In connection with the latter, other properties such as one-way propagation and slow light are also discussed.

This paper is summarized as follows: in Sec. II we present our conventions. Sec.\ III is devoted to some geometric properties of optical metrics. In Sec. IV, we restrict to the case of null backgrounds. In Sec. V, we introduce the Kerr-Schild form of the optical metric. Then, in Sec. VI we study some interesting examples. Finally, we discuss issues related to causality, energy conditions, and hyperbolicity in Sec. VII. Throughout the text, we use natural units.

\section{Conventions}
To begin with, we let $(M,g_{ab})$ denote a four-dimensional Minkowski spacetime with signature convention $(+,-,-,-)$. In an arbitrary coordinate system $\{x^{a}\}$, the electromagnetic field tensor and its Hodge dual are expanded, respectively, as
\begin{equation}
\boldsymbol{F}=\frac{1}{2}F_{ab}\ dx^{a}\wedge dx^{b},\quad\quad\quad \star\boldsymbol{F}=\frac{1}{2}\star F_{ab}\ dx^{a}\wedge dx^{b},
\end{equation}
where
\begin{equation}
\star F_{ab}=\frac{1}{2}\varepsilon_{ab}^{\phantom a\phantom a pq}F_{pq},
\end{equation}
with the Levi-Civita tensor given by $\varepsilon_{abcd}=\sqrt{-g}\ [abcd]$ and the permutation symbol defined such that $[0123]\equiv +1$. As usual, a future-directed, normalized, field of observers, say $v^{a}(x^{b})$, uniquely decomposes the field tensor as \footnote{Throughout, we stick to the conventions $(ab)=ab+ba$ and $[ab]=ab-ba$.}
\begin{equation}\label{dec}
F^{ab}=E^{[a}v^{b]}+\varepsilon^{ab}_{\phantom a\phantom a pq}B^{p}v^{q},\quad\quad\quad v^{q}v_{q}=1,
\end{equation}
with the electric and magnetic 3-vectors measured by the observer defined, respectively, by the projections
\begin{equation}
E^{a}=F^{a}_{\phantom a b}v^{b},\quad\quad\quad B^{a}=-\star F^{a}_{\phantom a b}v^{b}.
\end{equation}
From the latter, it is convenient to define another 3-vector (the Poynting vector)
\begin{equation}\label{Poynt}
S^{a}=\varepsilon^{abcd}v_{b}E_{c}B_{d},
\end{equation}
which will play a prominent role in our subsequent discussion. For the sake of conciseness, we define also the norm of an arbitrary 3-vector $X^{a}$ as
\begin{equation}
X\equiv\sqrt{-X^{a}X_{a}},
\end{equation}
and from the above definitions, there follows the orthogonality relations
\begin{equation}
e_{a}v^{a}=0,\quad\quad\quad b_{a}v^{a}=0,\quad\quad\quad s_{a}v^{a}=0,\quad\quad\quad e_{a}s^{a}=0,\quad\quad\quad b_{a}s^{a}=0,
\end{equation}
with the triad $\{\boldsymbol{e},\boldsymbol{b},\boldsymbol{s}\}$ denoting the corresponding unitary 3-vectors parallel to the electric field, magnetic field, and Poynting vector, respectively.

With the above conventions, the two independent invariant scalars of the electromagnetic field are defined by the relations
\begin{eqnarray}\label{invariants}
&&\phi\equiv \frac{1}{2}\star F_{ab}F^{ab}=2EB\mbox{cos}\ \theta,\quad\quad\quad \psi\equiv \frac{1}{2}F_{ab}F^{ab}=B^{2}-E^{2},
\end{eqnarray}
where $\theta$ is the angle between the electric and the magnetic 3-vectors as measured by $v^{q}$. At a spacetime point $p\in M$, the algebraic type of the field tensor is related to the number of independent solutions of the eigenvalue/eigenvector problems
\begin{equation}
k_{[a}F_{b]c}k^{c}=0,\quad\quad\quad k_{[a}\star F_{b]c}k^{c}=0.
\end{equation}
A lightlike eigenvector $k^{a}$ of this problem is called a principal null direction (PND) of the field. There follow two mutually exclusive possibilities: i) when the invariants vanish simultaneously, the field is called \textit{null} (or algebraically special) and there exists a single degenerate PND; ii) when at least one invariant is nonzero the field is called \textit{regular} (or algebraically general) and there exist two independent PNDs.

\section{Optical metric}

From now on, we shall be concerned with a gauge-invariant nonlinear theory of electrodynamics in $(M, g_{ab})$ provided by the action \footnote{The inclusion of the invariant $\phi$ does not affect the net result of the paper.}
\begin{equation}\label{action}
S=\int \mathcal{L}(\psi)\sqrt{-g}\ d^{4}x,
\end{equation}
where the Lagrangian density $\mathcal{L}(\psi)$ is, up to now, an arbitrary function of the invariant and is assumed to be well defined for $\psi=0$ in order to guarantee the existence of plane wave solutions. Writing $F_{ab}=\partial_{[a}A_{b]}$ and varying with respect to the four-potential yields the nonlinear extension of Maxwell electrodynamics in vacuum
\begin{equation}\label{nled}
\left(\mathcal{L}_{\psi}F^{ab}\right)_{;b}=0,\quad\quad\quad \left(\star F^{ab}\right)_{;b}=0,
\end{equation}
with $\mathcal{L}_{\psi}\equiv \partial\mathcal{L}/\partial\psi$ for conciseness and $;$ standing for covariant derivative with respect to the metric $g_{ab}$. A necessary (but not sufficient) condition for this theory to admit a well-posed Cauchy problem is that $\mathcal{L}_{\psi}\neq 0$, which we shall assume henceforth. This means that we solely consider Lagrangian densities that are strictly monotonic for all possible values of the invariant (see \cite{Abalos:2015gha, Goulart:2021uzr} for issues related to covariant hyperbolizations).

A direct consequence of Eqs. (\ref{nled}) is the automatic violation of the principle of superposition. In turn, this implies that linearized wavy disturbances about a smooth background solution propagate nontrivially. Borrowing from the terminology of laser optics, we may say that \textit{probe fields}
(small amplitude/high-frequency waves) interact with \textit{pump fields} (background solutions) and are scattered by the latter. In this vein, a multitude of effects on the polarization, wave-covector, frequency, and velocity of the ‘photons’ that probe the pump field may take place. Indeed, by perturbing the quasi-linear system of first-order PDEs Eqs. (\ref{nled}) around a fixed background solution and taking the eikonal limit for the excitation, we obtain an optical metric of the form
\begin{equation}\label{eq:cov_metric}
\tilde{g}_{ab}=g_{ab}-\frac{\kappa}{1-\kappa\psi} F_{a c}F^{c}{}_{b},
\end{equation}
with inverse given by
\begin{equation}\label{eq:contra}
\tilde{g}^{ab}=g^{ab}+\kappa F^{a}_{\phantom a c}F^{cb}.
\end{equation}
Here, the quantity $\kappa\equiv-2\mathcal{L}_{\psi\psi}/\mathcal{L}_{\psi}$ controls the magnitude of the nonlinearity whereas $F_{ab}$ describes the specific structure of the pump field. Roughly speaking, this means that the disturbances of the theory in the geometric optics regime do not travel along Minkowskian null lines. Rather, they are described by the null geodesics of an \textit{optical metric}, which depends implicitly on the background solution. In other words, the light rays $\xi^{a}(x^{b})$ must satisfy the coupled system of equations
\begin{equation}\label{disprel}
\tilde{g}_{ab}\xi^{a}\xi^{b}=0,
\end{equation}
\begin{equation}\label{disprel2}
\xi^{a}_{\phantom a||b}\xi^{b}=0,
\end{equation}
with $\tilde{g}^{ac}\tilde{g}_{cb}=\delta^{a}_{\phantom a b}$ and $||$ the covariant derivative compatible with $\tilde{g}_{ab}$. In general, this optical metric will not be flat. Indeed, introducing the Christoffel symbols as usual, we recall that for any pair of non-degenerate symmetric tensors $g_{ab}$ and $\tilde{g}_{ab}$, there follows the tensorial relation
\begin{equation}\label{C}
C^{a}{}_{bc}\equiv\tilde{\Gamma}^{a}{}_{bc} - \Gamma^{a}{}_{bc} = \frac{1}{2}\tilde{g}^{ad}(\tilde{g}_{db;c}+\tilde{g}_{cd;b}-\tilde{g}_{bc;d}).
\end{equation}
Furthermore, since we are dealing with a base Minkowski spacetime, it follows that the curvature tensor of the optical metric reads as
\begin{equation}\label{R}
\tilde{R}^{a}{}_{bcd}=C^{a}{}_{b[c;d]}+C^{a}{}_{e[d}C^{e}{}_{bc]}.
\end{equation}

\section{Null backgrounds}

Henceforth, we shall be concerned solely with pump fields of the null type. The main reason behind our assumption is that it engenders an optical metric with simple algebraic and differential structures which are well known to relativists. Furthermore, since a regular background is endowed with two independent PNDs, it seems that the corresponding optical metric would not admit the existence of the wave trappings, as discussed in Section VI. Therefore, we consider the exact solutions of Eqs. (\ref{nled}) such that the null constraints
\begin{equation}\label{constr}
\phi=0\quad\mapsto\quad\boldsymbol{e}\perp\boldsymbol{b},\quad\quad\quad \psi=0\quad\mapsto\quad E=B,
\end{equation}
hold in a region $\mathcal{U}\subseteq M$. A curious consequence of these assumptions is that the nonlinear equations of motion somehow linearize and, therefore, reduce to Maxwell equations in vacuum. Therefore, the background solution will be given by simple bivectors such that
\begin{equation}\label{Max}
F^{ab}_{\phantom a\phantom a;b}=0,\quad\quad\quad \star F^{ab}_{\phantom a\phantom a;b}=0.
\end{equation}
Examples of null solutions include plane and spherical waves, as well as the celebrated electromagnetic knots discussed in \cite{arrayas2017knots, irvine2008linked, kedia2013tying}. Also important is the fact that the retarded electromagnetic field from an isolated extended source has the following asymptotic behavior (property of sequential degeneration)
\begin{equation}
F_{ab}=\frac{F_{ab}^{(1)}}{r}+\frac{F_{ab}^{(2)}}{r^{2}}+\mathcal{O}(r^{-3}),
\end{equation}
where $r$ is an affine parameter along a principal null congruence of the field $F_{ab}$ and the fields $F_{ab}^{(1)}$ and $F_{ab}^{(2)}$ are independent of $r$ and of null and regular types, respectively. This result, also known as the Goldberg-Kerr theorem, shows that null fields play a crucial role in understanding the behavior of radiation as one goes to infinity.

From the algebraic perspective, the null constraints above permit us to recast the $(3+1)$ decomposition Eq.\ (\ref{dec}) in the reduced form
\begin{equation}\label{nullity}
F^{ab}=E^{[a}l^{b]},\quad\quad\quad \star F^{ab}=-B^{[a}l^{b]},
\end{equation}
with $E^{a}$ and $B^{a}$ being the electric and magnetic fields measured by the observer $v^{a}$ as before and $l^{a}$ the unique PND (up to rescalings) associated to the null field. It is easy to check that this PND is given by
\begin{equation}
l^{a}\equiv v^{a}+s^{a}.
\end{equation}
Consequently, there follows the relations
\begin{equation}
l^{a}l_{a}=0,\quad\quad l^{a}v_{a}=1,\quad\quad l^{a}s_{a}=-1\quad\quad l^{a}e_{a}=0.
\end{equation}
We shall see next that these relations drastically simplify the form of the optical metric. However, before we proceed, it is worth recalling that the kinematics of a null congruence $l^{a}(x^{b})$ in Minkowski spacetime is partially characterized by the acceleration vector
\begin{equation}
a^{p}=l^{p}_{\phantom a;q}l^{q}
\end{equation}
and the following optical scalars
\begin{equation}
\theta=\frac{1}{2}l^{p}_{\phantom a;p},\quad\quad\sigma=\left[\tfrac{1}{4}l_{(p;q)}l^{p;q}-\theta^{2}\right]^{1/2},\quad\quad\omega=\left[ \tfrac{1}{4}l_{[p;q]}l^{p;q}\right]^{1/2},
\end{equation}
respectively called the divergence, the shear and the twist scalars. Physically, these quantities are related to the dimension, shape, and orientation of the shadow projected by an opaque object onto a close screen, as guaranteed by Ehlers-Sachs theorem \cite{sachs1961gravitational}.  Furthermore, an important result originally due to Mariot \cite{mariot1954champ} and Robinson \cite{robinson1961null} states that the principal lightlike congruence of an algebraically special solution of the free Maxwell equations is necessarily geodesic ($a^{p}=0$) and shear-free ($\sigma=0$). We refer the reader to \cite{frolov1979newman} for a direct proof of this assertion in the context of the Newman-Penrose formalism. In the next section, we briefly discuss some aspects of the optical metrics constructed with these geodesic and shear-free null congruences.

\section{Kerr-Schild-like metrics}

Using the null field ansatz described in the last section in Eqs.\  (\ref{eq:cov_metric}) and (\ref{eq:contra}), we obtain the following covariant and contravariant optical metrics
\begin{eqnarray}
\tilde{g}_{ab}&=&g_{ab}- H l_{a}l_{b},\label{eff1}\\[1ex]
\tilde{g}^{ab}&=&g^{ab}+ H l^{a}l^{b},\label{eff2}
\end{eqnarray}
with the scalar function given by $H\equiv \kappa E^{2}$, for conciseness. Since the derivatives of the lagrangian are to be evaluated on top of $\psi=0$ and $\phi=0$, they are constants, and the behavior of $H$ is essentially governed by the squared norm of the electric field, which is somehow constrained to satisfy Maxwell equations in vacuum. Furthermore, since the PND $l^{a}$ must be tangent to a geodesic shear-free congruence in $(M,g_{ab})$, the optical metric associated to the null background is remarkably similar to the Kerr-Schild metrics, which are well known in the context of black hole physics. In particular, the Ricci tensor associated to Eq. (\ref{R}) is given by \cite{Stephani_2003}
\begin{equation}
\tilde{R}_{ab} =  \frac{1}{2}[(H l_a l_b)^{;c}{}_{;c} - (H l^c l_b)_{;ca} - (H l^c l_a)_{;cb} + H H''l_a l_ b ],
\end{equation}
with $H'\equiv H_{;a}l^{a}$. A straightforward calculation then shows that the corresponding Ricci scalar vanishes identically. Before we proceed to the examples, let us briefly summarize some known aspects of Kerr-Schild metrics. If we think of Eq.\ (\ref{eff2}) as some kind of deformation of the flat spacetime metric, we have the following properties:
\begin{enumerate}

\item{It preserves the null character of the PND. Furthermore, the PND is the unique null vector (up to rescallings) satisfying this property;}

\item{The PND is geodesic and shear-free also in the manifold $(M,\tilde{g}_{ab})$;}

\item{It is a volume-preserving transformation since $\mbox{det}(\tilde{g}_{ab})=\mbox{det}(g_{ab})$. In other words, the expansion coefficient (divergence of $l^a$) is the same for both metrics. The same holds for the twists};

\item{The PND is a principal null direction of the optical Riemann tensor. Indeed, there follows $\tilde{R}_{abcd}l^{b}l^{d}\sim H''l_{a}l_{c}$. Consequently, it is also a principal null direction of the Weyl tensor, indicating that this space-time is algebraically special. In particular, the Newman-Penrose (NP) curvature invariants $\Psi_0$ and $\Psi_1$ vanish.}

\end{enumerate}

\noindent Besides the above properties, the most important feature of Eq. (\ref{eff1}) to our analysis is that the induced ray equation Eq. (\ref{disprel}) admits the following solutions in the 2-plane spanned by the observer and the Poynting vector:

\begin{equation}
l^{a}=v^{a}+s^{a}, \qquad \mbox{and} \qquad n^{a}=v^{a}+Ys^{a},
\end{equation}
where $Y\equiv (H - 1)/(H + 1)$. This means that signals emitted along the direction of the Poynting vector will travel with the velocity of ordinary light and signals emitted contrary to this direction will travel with another velocity, given by $Y$. From now on, we shall stick to the following terminology: $l^{a}$ will be called the \textit{forward ray} and $n^{a}$ the \textit{backward ray}. Routine calculations then show that the backward ray will be timelike with respect to $g_{ab}$ whenever $\kappa> 0$, which we now assume. It is straightforward to show that if the latter holds, all rays (except those proportional to $l^{a}$) will also be timelike with respect to $g_{ab}$.

\begin{figure}[ht]
    \centering
    \vspace{-2cm}
    \includegraphics[scale=.4]{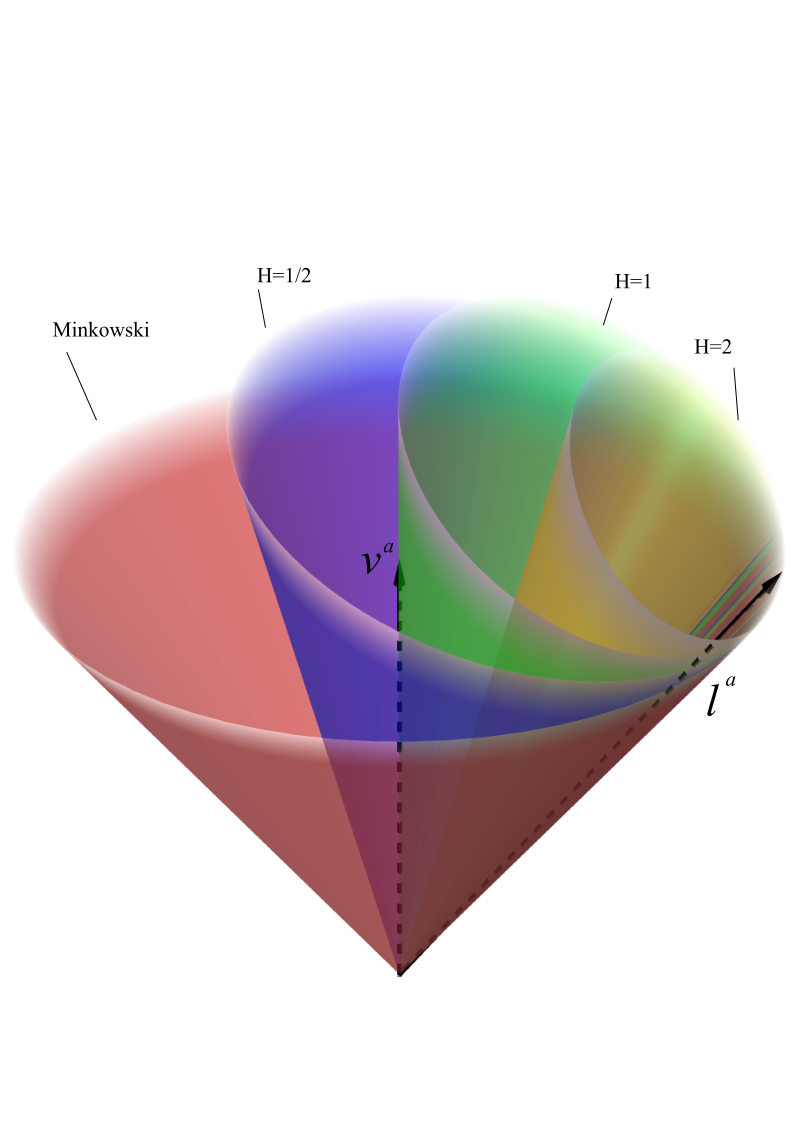}
    \vspace{-1.5cm}
    \caption{Color online. Geometric picture of the optical cones compared to the Minkowski light cone (red). The values of the parameter $H$ were chosen to be $1/2$ to represent the sub-critical case (blue), $1$ for the critical case (green), and $2$ to represent the super-critical case (yellow). Note that all cones share the same principal null direction $l^a$.}
    \label{fig:cones}
\end{figure}

A drastic consequence of our discussion is that for large enough values of the electric/magnetic field, the backward ray $n^{a}$ will be at rest with respect to the observer $v^{a}$. This effect happens whenever the scalar function $Y$ vanishes i.e.
\begin{equation}
\label{eq:crit_case}
H - 1=0\quad \longleftrightarrow\quad E^{2}=1/\kappa.
\end{equation}
From this, we can divide the range of interest of $H$ into three intervals: $0\leq H<1$ sub-critical, $H=1$ is critical, and $H>1 $ super-critical. The qualitative behavior of these cases is depicted in Fig.\ (\ref{fig:cones}). A time-like observer at rest with respect to the Minkowski light cone is also time-like with respect to the sub-critical optical cone. In the critical case, the observer is comoving with the backward ray. For the super-critical optical cone, the observer lies outside the causal region of the theory. Perhaps not surprisingly, these situations have a close analogy with aircrafts traveling in the subsonic, transonic, and supersonic regimes. 

\section{Examples}

In this section, we shall demonstrate the existence of three possible photon traps within the framework presented before. By a photon trap, we mean a scheme for confining the photon (excitation) in a region of spacetime such that a distant observer cannot detect it. Furthermore, the examples provided here are quite universal in the sense that the only hypothesis underlying them are 

\begin{itemize}
\item{a reasonable null background solution;}
\item{a strictly positive parameter $\kappa$;}
\item{a sufficiently high electromagnetic intensity;}

\end{itemize}
\subsection{Constant field}

To start with, write $x^{a}=(t,x,y,z)$ for Cartesian coordinates in Minkowski spacetime and consider the following open slab
\begin{equation}
\mathcal{U}=\{x^{a}\in M| -z_{0}<z<z_{0}\},
\end{equation}
with $z_{0}$ any positive constant. Letting $v^{a}=(1,0,0,0)$ denote a field of inertial observers with trivial kinematics, we shall assume that there exists a background electromagnetic field in $\mathcal{U}\subseteq M$, such that:
\begin{equation}
E^{a}=(0,E_{0},0,0),\quad\quad B^{a}=(0,0,E_{0},0),\quad\quad l^{a}=(1,0,0,1),
\end{equation}
with $E_{0}$ also a positive constant. In order to avoid possible discontinuity problems, assume that outside this region the fields vary smoothly until they vanish. In principle, this simple \textit{ansatz} could be achieved by embedding a large capacitor inside an ideal solenoid and adjusting the field directions and intensities appropriately. Clearly, this configuration generates a null solution of Maxwell equations in vacuum Eqs. (\ref{Max}), and the optical line element constructed with  Eq.\ \eqref{eff1} reads simply as
\begin{equation}
\label{eq:opt_line_const_case}
    d\tilde{s}^{2}=\left(1-\kappa E_{0}^{2}\right)dt^{2}-dx^{2}-dy^{2}-\left(1+\kappa E_{0}^{2}\right)dz^{2}+ 2\kappa E_{0}^{2}dtdz.
\end{equation}
Since all metric coefficients above are constant, the corresponding null geodesics are still straight lines. However, one easily sees that the corresponding light cones are somehow dragged towards the direction of the Poynting vector, very much in the same way as a moving fluid will drag sound waves along with it. In particular, if we increase the intensity of the electric field in order to achieve the critical regime, the optical line element becomes
\begin{equation}
\label{eq:opt_line_const_case_crit}
    d\tilde{s}^{2}=2dtdz-dx^{2}-dy^{2}-2dz^{2}.
\end{equation}
We notice that $z$ now becomes a null coordinate, meaning that each hyperplane of constant z inside the slab separates the space-time into two disjoint regions and the corresponding one-form $dz$ is a null direction. Therefore, the hyperplanes work as one-way membranes for the light rays: the latter can travel from the region $z<z_{0}$ to the region $z>z_{0}$ without problems, but not in the opposite direction. A helpful way of thinking of this case is to imagine that someone turns on a lantern at any point in the region $z>z_{0}$ and, as a result, only half of the room gets bright. Needless to say, if we increase the magnitude of the electric field any further, $H$ enters the super-critical regime and light rays get even more confined in the room: the slab works as a reflecting mirror!

\begin{figure}
    \centering
    \includegraphics[scale=.41]{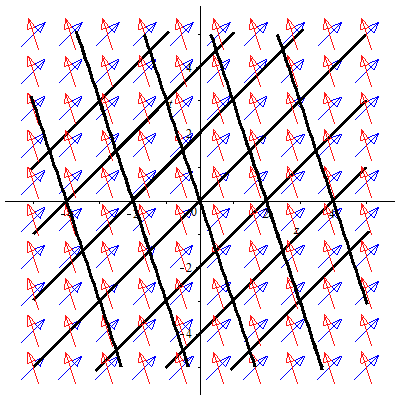}
    \includegraphics[scale=.41]{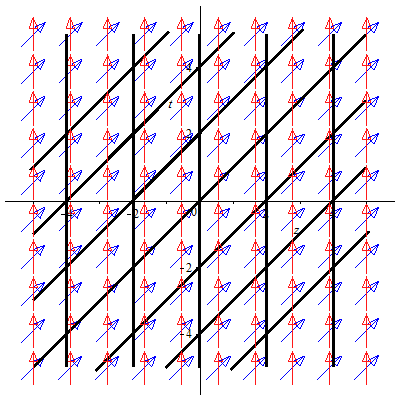}
    \includegraphics[scale=.41]{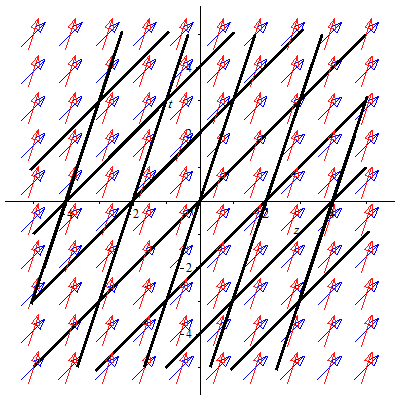}
    \caption{Spacetime diagram for the constant field case. Left. $\kappa E_0^2=1/2$. Center. $\kappa E_0^2=1$. Right. $\kappa E_0^2=2$.}
    \label{fig:cones_constant_field}
\end{figure}

\subsection{Plane wave}

Consider a Cartesian coordinate system as before and a set of two covariantly constant vector fields $e^{a}$ and $l^{a}$, say. Defining the phase function as $\Phi\equiv l_{a}x^{a}$, we construct the following simple bivector
\begin{equation}\label{bivec}
F^{ab}=h(\Phi)\big(e^{a}l^{b}-e^{b}l^{a}\big),
\end{equation}
with $h(\Phi)$ an arbitrary function of the phase. From the latter we get
\begin{equation}
F^{ab}_{\phantom a \phantom a;c}=h_{\Phi}e^{[a}l^{b]}l_{c}\quad\quad\quad \star F^{ab}_{\phantom a \phantom a;c}= h_{\Phi}\varepsilon^{ab}_{\phantom a\phantom a pq}e^{p}l^{q}l_{c},
\end{equation}
with $h_{\Phi}$ denoting derivative with respect to $\Phi$. A straightforward calculation then shows that Eq. (\ref{bivec}) will be a null solution of Maxwell's equations in vacuum Eqs. (\ref{Max}), provided we have:
\begin{equation}
e^{a}e_{a}=-1,\quad\quad l^{a}l_{a}=0,\quad\quad l^{a}e_{a}=0.
\end{equation}
Clearly, this solution describes a plane-fronted wave with principal null direction $l^{a}$ (which is obviously geodesic shear-free) and linear polarization vector $e^{a}$. Without loss of generality, we assume that the wave is traveling in the $+z$ direction i.e.,
\begin{equation}
e^{a}=(0,1,0,0), \quad\quad b^{a}=(0,0,1,0),\quad\quad  s^{a}=(0,0,0,1).
\end{equation}
A direct inspection of Eq. (\ref{eff1}) then shows that the corresponding optical line element will be quite similar to the one of the constant case: all one has to do is to make the substitution $E_{0}\mapsto h(\Phi)$ in Eq. (\ref{eq:opt_line_const_case}) and the crucial difference here is that the optical line element now varies in time and space. Hence, the dragging effect on the light cones will not be homogeneous as in the previous case but will depend on the intensity of the pump field on the given event.

\begin{figure}
    \centering
    \includegraphics[scale=.41]{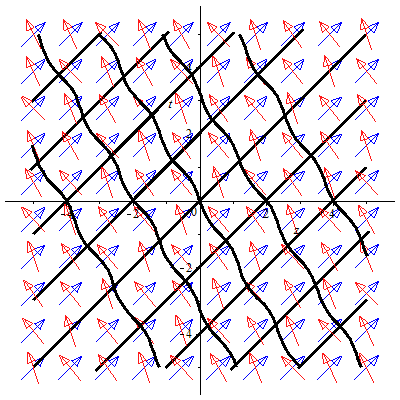}
    \includegraphics[scale=.41]{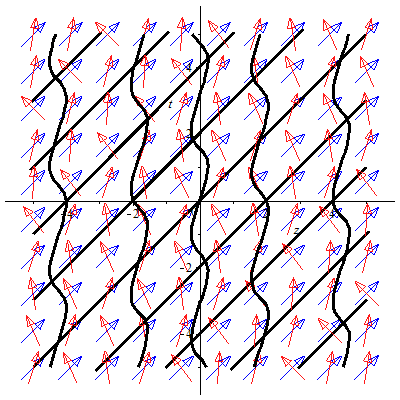}
    \includegraphics[scale=.41]{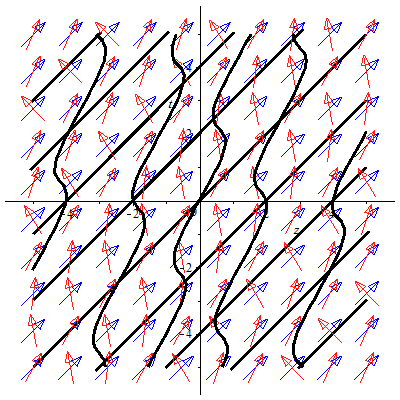}
    \caption{Spacetime diagram for the plane wave case. Left. $\kappa E_0^2=0.5$. Center. $\kappa E_0^2=2.0$. Right. $\kappa E_0^2=3.5$. Black straight lines tilted to the right are the integral lines of the null vector $l^a$, while the periodic curves tilted to the left are the integral curves of the vector field $n^a$.}
    \label{fig:cones_planes_waves}
\end{figure}

Is the optical metric above curved? A straightforward calculation using Eq. (\ref{C}) reveals that the Christoffel symbols read as
\begin{equation}
\tilde\Gamma^{a}_{bc}=H_{\Phi}l^a l_b l_c.
\end{equation}
Although they are not constant, it is easy to show that the corresponding Riemann tensor vanishes. Therefore, we are again dealing with a flat optical metric. Interestingly enough, the behavior of the causal structure is quite different from the constant case. For the sake of illustration, imagine a low amplitude X-ray propagating on top of a high intensity infrared pump field. Choosing a \textit{normalized} sinusoidal function $h(\Phi)$ to represent the shape of the latter, we have 
$$H=\kappa E_0^2\cos^2(t-z),$$
which implies in the inequality $0\leq H\leq \kappa E_0^2$. Then, a direct computation reveals that in the plane $t\times z$ the forward and backward rays must satisfy the ordinary differential equation
\begin{equation}\label{forback}
\left[1+\kappa E_{0}^{2}\cos^2(t-z)\right]\left(\frac{dz}{dt}\right)^2 - 2\kappa E_{0}^{2}\cos^2(t-z)\frac{dz}{dt}-\left[1-\kappa E_{0}^{2}\cos^2(t-z)\right]=0.
\end{equation}
Therefore, the two independent solutions of Eq. (\ref{forback}) are
\begin{equation}
\frac{dz_+}{dt}=1,\quad\quad\quad \frac{dz_-}{dt}=\frac{\kappa E_{0}^{2}\cos^2(t-z)-1}{\kappa E_{0}^{2}\cos^2(t-z)+1},
\end{equation}
The latter shows that the forward rays indeed coincide with Minkowskian null lines, whereas the backward rays may be implicitly obtained in terms of the level sets of the function 
\begin{equation}
G(t,z)=(\kappa E_{0}^{2}+2)z_{-}-(\kappa E_{0}^{2} - 2) t - \frac{\kappa E_{0}^{2}}{2}\sin[2(t - z_{-})].
\end{equation}
A careful inspection of the latter shows that the trajectories somehow oscillate in spacetime and are generally biased towards a direction, depending on the field intensity. When $\kappa E_0^2<2$, the net result of the oscillations is towards left whereas when $\kappa E_0^2>2$ the result is towards right. Interestingly, for the particular field intensity $\kappa E_0^2=2$ there is a quite unexpected regime: $G(t,z)$ becomes a periodic function in time with time periodicity given by $t_{0}=\pi$. This means that a inertial observer located at the position $(x_{0},y_{0},z_{0})$ effectively  \textit{sees} the backward ray as an oscillating photon (see Fig.\ \ref{fig:cones_planes_waves}).

\subsection{Dipolar wave}

Let $x^{a}=(t,r,\theta,\phi)$ be a spherical coordinate system in $(M,g_{ab})$ and consider the geodesic shear-free null congruence defined by the vector field $l^{a}=(1,1,0,0)$. Defining the phase function $\Phi$ as before and the positive constant $p_{0}$, it can be checked that the bivector field
\begin{equation}\label{dipole}
F^{ab}= p_{0}\left(\frac{\mbox{sin}\theta}{r^{2}}\right)\mbox{cos}(\Phi)\left(\begin{array}{cccc}
\ 0&\ 0&1 &0\\
\ 0&\ 0&1 &0\\
-1 &-1 &0 &0 \\
\ 0&\ 0&0 &0\\
\end{array}\right)
\end{equation}
is algebraically special (null), with $l^{a}$ being the corresponding PND. Eq.\ (\ref{dipole}) represents a monochromatic wave traveling in the radial direction and corresponds to the field generated by an oscillating electric dipole in the radiation zone (see Ref.\ \cite{jackson2012classical} for more details). In other words,  it is a valid solution of Eqs.\ (\ref{Max}) as far as we have 
\begin{equation}
d \ll \lambda\ll r,
\end{equation}
where $d$ is the dipole separation and $\lambda$ is the wavelength. We notice that the electric field is tangent to the meridian lines whereas the magnetic field is purely azimuthal, thus yielding a Poynting vector orthogonal to spheres of constant radii. However, there is a dependence on $\theta$ which implies that there is no radiation along the axis of the dipole: the intensity profile takes the form of a donut, with its maximum in the equatorial plane.

\begin{figure}
    \centering
    \includegraphics[scale=.42]{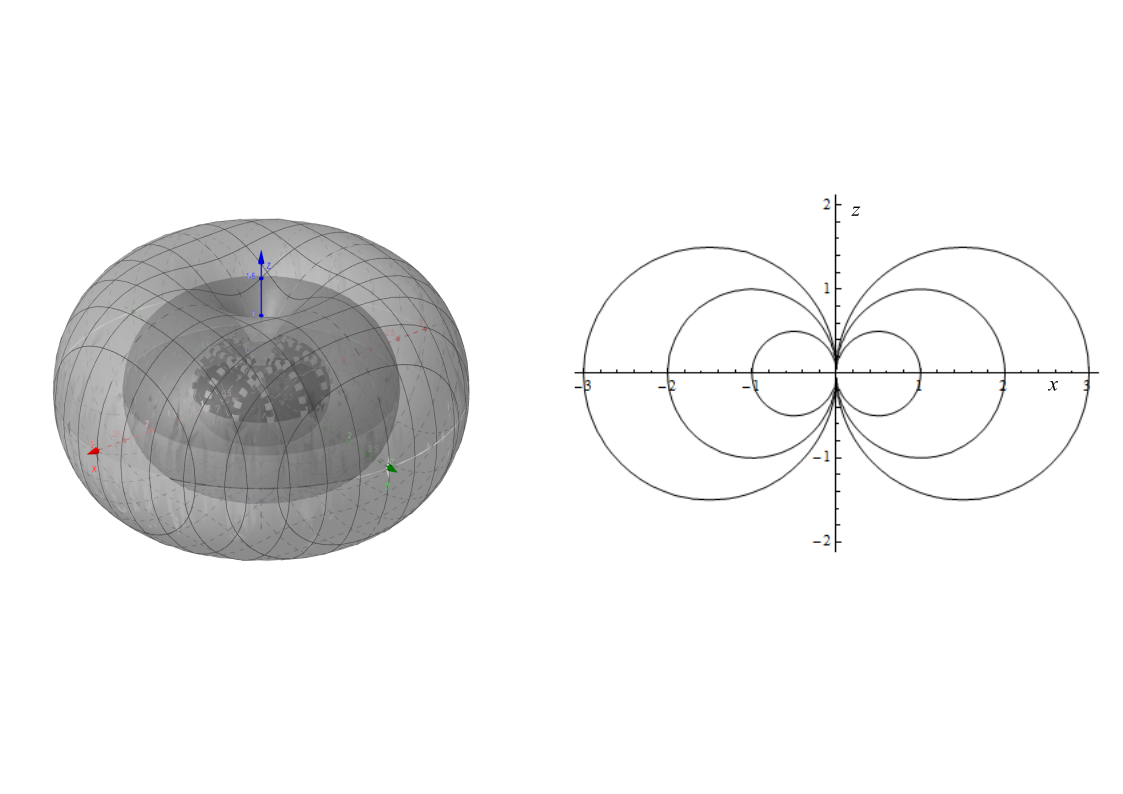}
    \vspace{-2cm}
    \caption{Left. The 3-dimensional profile of the averaged electric field lines of a dipole. Right. The section of some field lines in the (x,z)-plane. The azimuthal symmetry indicates that they are horn tori.}
    \label{fig:equip_dipole}
\end{figure}

The (approximate) solution we are interested in can be obtained from Eq.\ (\ref{dipole}) by simultaneously changing the signs of magnetic field and Poynting vector, i.e.
\begin{equation}
\boldsymbol{b}\ \rightarrow\ -\boldsymbol{b},\quad\quad\quad \boldsymbol{s}\ \rightarrow\ -\boldsymbol{s}.
\end{equation}
This represents the inverse situation: a spherical collapse of monochromatic electromagnetic waves. Clearly, the effective metric constructed with this solution is time-dependent. However, in order to simplify our analysis we proceed very much in the same way as textbooks on radiation theory. In the case of visible light, for instance, the wavelength is so short and the period so brief that any macroscopic measurement will encompass many cycles. Typically, therefore, we are not interested in the fluctuating cosine-squared: all we want is the average over a complete cycle, i.e.,
\begin{equation}
\langle E^{2}\rangle = \frac{p_{0}^{2}}{2}\left(\frac{\mbox{sin}^{2}\theta}{r^{2}}\right).
\end{equation}
The profile of the mean electric field is depicted in Fig.\ (\ref{fig:equip_dipole}). Technically, the surfaces correspond to horn tori. 

With these assumptions, the \textit{averaged} optical line element reads as
\begin{equation}\label{eq:opt_line_sph_case}
   d\tilde{s}^{2}=\left(1-\frac{M_{0}\sin^{2}\theta}{r^{2}}\right)dt^{2}-\left(1+\frac{M_{0}\sin^{2}\theta}{r^{2}}\right)dr^{2} - \frac{2M_{0}\sin^{2}\theta}{r^{2}} dt\,dr -r^{2}d\Omega^{2},
   \end{equation}
with $M_{0}\equiv \kappa p_{0}^2/2$ and $d\Omega$ the usual line element on the unit sphere. Clearly, at sufficiently large distances, this optical geometry becomes indistinguishable from that of Minkowski spacetime. Furthermore, we see that the same situation holds along the $z$ axis since the electromagnetic field vanishes there. Note also the resemblance between this line element and the Schwarzschild metric in Eddington-Finkelstein coordinates \cite{gravitation2017}. Introducing the coordinate transformation $\chi=\sin\theta$ to simplify the calculations \cite{visser2008kerr}, commonly called rational polynomial coordinate, we can easily conclude, using a symbolic manipulation package, that all invariants are well behaved, with the possible exception of the origin. In particular, the Kretschmann invariant has the simple form
\begin{equation}
K=R_{abcd}R^{abcd}=\frac{56 M_0^2\chi^4}{r^8},
\end{equation}
and blows up for $r\rightarrow 0$, showing that there is a real singularity there. Furthermore, since $\chi$ vanishes for $\theta=0$, this reinforces the fact that the optical spacetime is indeed flat along this direction. A straightforward calculation gives that this metric is Petrov type II.

\begin{figure}
    \centering
    \includegraphics[scale=.42]{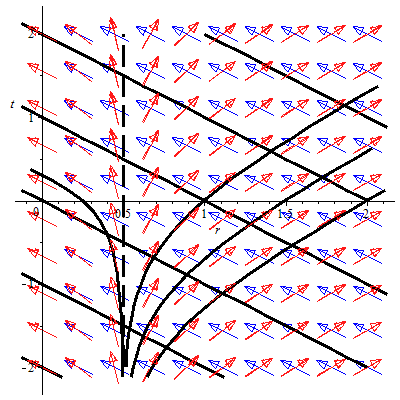}
    \includegraphics[scale=.42]{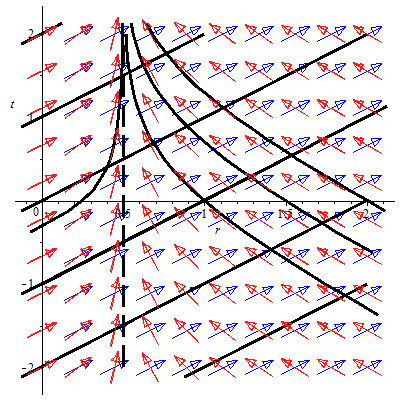}
    \caption{Color online. Phase portraits generated by $l^a$ (blue) and $n^a$ (red) are restricted to the 2D spacetime diagram (t,r), with $\theta=\pi/2$. Solid black lines are integral curves of the field $n^a$ and the vertical dash line indicates the location of a one-way membrane. Left. Case of an inward Poynting vector. Right. Case of an outward Poynting vector.}
    \label{fig:st_diagram_spherical}
\end{figure}

For this case, the backward ray is $n^a=(1,Y,0,0)$, where
\begin{equation}
Y=\fracc{M_0\sin^2\theta - r^2}{M_0\sin^2\theta + r^2}.
\end{equation}
The phase portraits generated by $l^a$ (blue) and $n^a$ (red) are restricted to the 2D spacetime diagram spanned by (t,r), with $\theta=\pi/2$, and they are depicted in Fig.\ (\ref{fig:st_diagram_spherical}). The solid black lines are integral curves of the field $n^a$, while the vertical dash line indicates the location of a one-way membrane (here it is a circle). On the figure left-hand side, we have the case of an inward Poynting vector. Thus, the membrane confines the light rays and does not let them escape outside. On the right-hand side, the Poynting vector points outward and the membrane works in the opposite direction, and light rays cannot enter that region.

Now, we shall demonstrate that this metric admits special surfaces similar to the Kerr metric. First, the stationary character of the metric guarantees the existence of an infinite redshift surface for distant observers at rest, that is, this spacetime has the analogous of an ergosphere given by
\begin{equation}
    g_{00}=0\quad\Longrightarrow\quad r^{2}=M_0\sin^{2}\theta.
\end{equation} 
As we said before, this equation corresponds to a horn torus. And, as it happens for the Kerr metric, the ``stationary limit'' surface is not trapped. Indeed, its gradient is a spacelike vector except at the equatorial plane. In the attempt to find trapped surfaces, we assume the existence of an axially symmetric surface $\Sigma=\Sigma(r,\theta)$, and take its gradient $n_a=\partial_{a}\Sigma=(0,\frac{\partial\Sigma}{\partial r},\frac{\partial\Sigma}{\partial \theta},0)$. The condition for $\Sigma$ to be a marginally trapped surface is
\begin{equation}
    g^{ab}n_a n_b=0\quad\longrightarrow\quad \left(1-\frac{M_{0}\sin^{2}\theta}{r^{2}}\right)\left(\frac{\partial\Sigma}{\partial r}\right)^2+\frac{1}{r^2}\left(\frac{\partial\Sigma}{\partial \theta}\right)^2=0.
\end{equation}
Unfortunately, this is a non-linear PDE that cannot be solved using the usual techniques of handbooks \cite{polyanin}. However, on the equatorial plane, it is easy to verify that the ergosphere works as an event horizon. For $\theta\neq\pi/2$, we have no intuition about the existence of a such compact trapped surface.

For the sake of completeness, we construct the geodesic equations for test particles moving in the metric (\ref{eq:opt_line_sph_case}), as follows
\begin{eqnarray}
&&\ddot t  + \frac{M_0^{2} \sin^4\theta}{r^{5}}\,\dot t^2 + \frac{ 2M_0\sin^2\theta (M_0\sin^2\theta + r^{2} )}{r^{5}}\,\dot t \dot r -\frac{2 M_0 \sin\theta\cos\theta}{r^{2}}\, \dot t \dot\theta + \frac{M_0\sin^2\theta (M_0\sin^2\theta + 2 r^{2})}{r^{5}}\, \dot r^2 \nonumber\\[1ex] 
&&-\frac{2M_0\sin\theta\cos\theta}{r^{2}}\,\dot r \dot\theta - \frac{M_0 \sin^2\theta}{r}\, \dot\theta^2 - \frac{M_0\sin^4\theta}{r}\, \dot\varphi^2=0,\\[1ex]
&&\ddot r  - \frac{( M_0\sin^2\theta - r^{2}) M_0 \sin^2\theta}{r^{5}}\, \dot t^2 - \frac{2 M_0^{2} \sin^4\theta}{r^{5}}\, \dot t \dot r +  \frac {2M_0 \sin \theta  \cos\theta}{r^{2}}\, \dot t \dot\theta - \frac{(M_0 \sin^2\theta + r^{2}) M_0 \sin^2\theta}{r^{5}}\, \dot r^2\nonumber\\[1ex] 
&&+ \frac{2M_0 \sin\theta \cos\theta}{r^{2}}\, \dot r \dot\theta + \frac{( M_0 \sin^2\theta - r^{2})}{r}\, (\dot\theta^2 + \sin^2\theta\dot\varphi^2)=0,\\[1ex]
&&\ddot \theta - \frac{H_{0}\cos\theta \sin\theta}{r^{4}}\, (\dot t + \dot r)^2 + \frac{2}{r}\,\dot r \dot\theta - \cos\theta \sin\theta  \dot \varphi^{2}=0,\\[1ex]
&&\ddot\varphi + \frac{2}{r}\,\dot r \dot\varphi + 2\cot\theta \dot\theta \dot\varphi=0
\end{eqnarray}
For null geodesics, there are three immediate first integrals given by
\begin{eqnarray}
&&\left(1-\frac{M_0\sin^{2}\theta}{r^{2}}\right)\dot t^{2}-\left(1+\frac{H_{0}\sin^{2}\theta}{r^{2}}\right)\dot r^{2}-r^{2}(\dot\theta^2 +\sin^2\theta\dot\varphi^2) - \frac{2H_{0}\sin^{2}\theta}{r^{2}} \dot t \dot r=0,\\[1ex]
&&\left(1-\frac{H_{0}\sin^{2}\theta}{r^{2}}\right)\dot t- \frac{H_{0}\sin^{2}\theta}{r^{2}} \dot r=E,\\[1ex]
&&r^2\sin^2\theta\dot \varphi=L,
\end{eqnarray}
where $E$ and $L$ are constant related to the particle energy and angular momentum for observers at infinity.

The analysis can be simplified by setting the motion along the equatorial plane ($\theta=\pi/2$ and $\dot\theta=0$). Thus, the study reduces to a one-dimensional problem of a particle under an effective potential given by
\begin{equation}
E^2=\dot r^{2} + \left(1-\frac{M_0}{r^{2}}\right)\frac{L^2}{r^2}.
\end{equation}
Note that the effective potential vanishes for $r=M_0$, it has a unique critical value at $r_c=\sqrt{2M_0}$, and it is attractive for $r<r_c$ and repulsive otherwise. Still, in analogy to the Eddington-Finkelstein coordinates, we can define a tortoise coordinate, as follows
\begin{equation}
\frac{dr^*}{dr}=\left(1-\frac{M_0}{r^2}\right)^{-1}\quad\longrightarrow\quad r^*=r+\frac{1}{2}\ln\left|\frac{r-\sqrt{H}}{r+\sqrt{H}}\right|.
\end{equation}
From this, we define null coordinates of the form $v=t+r^*$ and $u=t-r^*$. In particular, in the case of radial ingoing null geodesics, which is adapted to the fact that the Poynting vector is pointing towards the origin, we have that the $r(\lambda)$ is defined for all values of the affine parameter $\lambda$, but $v\rightarrow-\infty$ as $\tau\rightarrow\sqrt{M_0}$. This means that $r=\sqrt{M_0}$ plays the role of an event horizon for null geodesics. Therefore, light rays traveling on the equatorial plane will be trapped if they cross the circle of radius $M_0$. The same reasoning can be done in the case of a Poynting vector directed outwards, but now it simulates the behavior of a white hole.

\section{Causality, energy conditions and hyperbolicity}

We have seen that the nonlinear theory will be causal (in the sense of Minkowski spacetime) on top of null background fields whenever $\kappa>0$. In other words, the photon will propagate \textit{inside} the usual light cone if we assume different signs for $\mathcal{L}_{\psi}$ and $\mathcal{L}_{\psi\psi}$ when evaluated for $\psi=\phi=0$. However, one might wonder whether the examples presented so far are still in conflict with other features such as the energy conditions and hyperbolicity. In order to show that this is not the case, we start by constructing the energy-momentum tensor associated to Eq. (\ref{action}). Variation with the background metric gives
\begin{equation}
T_{ab}=\frac{2}{\sqrt{-g}}\frac{\delta \left(\sqrt{-g}\mathcal{L}\right)}{\delta g^{ab}}=-2\mathcal{L}_{\psi}F_{a}^{\phantom a c}F_{cb}-\mathcal{L}g_{ab}.
\end{equation}
Usually, one assumes that the lagrangian density is an analytic function at $\psi=\phi=0$ and that in the limit of weak
electromagnetic fields, the action reduces to that of the Maxwell theory. Therefore, since we are interested in the behavior of the theory on top of null backgrounds, we write
\begin{equation}\label{tab}
T_{ab}=-2\mathcal{L}_{\psi}E^{2}l_{a}l_{b},
\end{equation}
with $\mathcal{L}(0)=0$, in order to mimic Maxwell theory as far as possible. The null energy condition, $T_{ab}k^{a}k^{b}\geq 0$ (where $k^{a}$ is a null vector $g_{ab}k^{a}k^{b}=0$), is satisfied provided $\mathcal{L}_{\psi}<0$. As is well known, violation of this condition would imply an unbound from below Hamiltonian and hence signifies the instability of the system. Similarly, assuming $\mathcal{L}_{\psi}<0$, it is easy to show that, for a null background, the weak, the dominant as well as the strong energy conditions are automatically satisfied. Therefore, in order to harmonize all basic energy conditions with the causality assumption, all we must demand is that
\begin{equation}
\mathcal{L}_{\psi}<0,\quad\quad\quad \mathcal{L}_{\psi\psi}>0.
\end{equation}
A key feature to elucidate about the partial differential equations governing NLED is whether they pose an initial value formulation. Well-posedness is at the roots of physics, for it amounts to the predictability power of the theory, asserting that solutions exist, are unique, and depend continuously on the initial data. Interestingly, for any NLED, necessary and sufficient conditions theories and fields must satisfy in order to have a well-posed initial value formulation is presented in \cite{Abalos:2015gha}. Such condition translates very nicely into geometrical terms: the system is symmetric hyperbolic, if and only if, the two convex sets arising from the dispersion relations (or, conversely, the characteristic surfaces) have a non-empty intersection. It turns out that this is precisely the situation presented in our examples. Therefore, the trapping mechanism does not violate causality, energy conditions, and hyperbolicity.  

\section{Concluding remarks}

We investigated the existence of photon traps within the context of NLED. Due to the nonlinearities, we discuss the appearance of an effective optical metric that belongs to the class of Kerr-Schild metrics in the case of null background fields. Moreover, the null vector is a principal null direction of the electromagnetic energy-momentum tensor, leading to several algebraic and geometric features for the optical metric.

Then, from the fact that null backgrounds are common solutions of any NLED and linear Maxwell's theory, we explore the optical metric of classical radiation fields found in any textbook of electromagnetism. In particular, we studied constant waves, where we already see that there are values of the free parameters of the theory for which photons traveling in a certain direction are at rest for the Minkowskian observers, although the optical metric is still flat. A similar behavior occurs in the case of plane waves, but now the photon is confined in a region such that for the observers it keeps bouncing forth and back. Finally, we imitate a gravitational collapse investigating the case of dipole radiation, where trapped regions seem to be present, but their complete characterization is still an open question. Then, we discuss aspects of causality, energy conditions, and hyperbolicity, emphasizing that whenever $\kappa>0$, the notion of propagation is safe, namely, instabilities cannot emerge since all conditions above are satisfied. 

It is worth mentioning that, within the realm of analogue gravity, the construction of trapping regions for photons was based on finding an optical metric admitting trapped surfaces. However, to the best of our knowledge, all proposals found in the literature so far suffer from mathematical pathologies related to issues such as lack of hyperbolicity, regularity of the solution on the whole manifold, and violation of energy conditions. In particular, we notice that it is not possible to construct an analog model for black holes using a charged point particle. Therefore, as an attempt to bypass the difficulties aforementioned, we have seen here that an effective trapped beam of photons can be obtained by just narrowing the optical cone relative to the background cone. From the theoretical point of view, this will be the case for any nonlinear electrodynamics as long as its Lagrangian satisfies the qualitative features developed in this manuscript.

\acknowledgments
EB would like to thank CNPq for the financial support (grant N.\ 305217/2022-4). EB also thanks Y.O. Souza for his comments on a previous version of this manuscript.


\bibliography{refs}

\end{document}